\documentclass[aps,prb,reprint,groupedaddress,floatfix,superscriptaddress,amsmath,showpacs,nobibnotes]{revtex4-1}
\usepackage{multirow,tikz}

\begin{document}
\title{Raman scattering study of large magnetoresistance semimetals $\text{TaAs}_2$ and $\text{NbAs}_{2}$}
\author{Feng\,Jin}
\affiliation{Department of Physics, Beijing Key Laboratory of Opto-Electronic Functional Materials and Micro-nano Devices, Renmin University of China, Beijing 100872, People's Republic of China}
\author{Xiaoli\,Ma}
\affiliation{Department of Physics, Beijing Key Laboratory of Opto-Electronic Functional Materials and Micro-nano Devices, Renmin University of China, Beijing 100872, People's Republic of China}
\author{Pengjie\,Guo}
\affiliation{Department of Physics, Beijing Key Laboratory of Opto-Electronic Functional Materials and Micro-nano Devices, Renmin University of China, Beijing 100872, People's Republic of China}
\author{Changjiang\,Yi}
\affiliation{Beijing National Laboratory for Condensed Matter Physics, Institute of Physics, Chinese Academy of Sciences, Beijing 100190, People's Republic of China}
\author{Le\,Wang}
\affiliation{Beijing National Laboratory for Condensed Matter Physics, Institute of Physics, Chinese Academy of Sciences, Beijing 100190, People's Republic of China}
\author{Yiyan\,Wang}
\affiliation{Department of Physics, Beijing Key Laboratory of Opto-Electronic Functional Materials and Micro-nano Devices, Renmin University of China, Beijing 100872, People's Republic of China}
\author{Qiaohe\,Yu}
\affiliation{Department of Physics, Beijing Key Laboratory of Opto-Electronic Functional Materials and Micro-nano Devices, Renmin University of China, Beijing 100872, People's Republic of China}
\author{Jieming\,Sheng}
\affiliation{Department of Physics, Beijing Key Laboratory of Opto-Electronic Functional Materials and Micro-nano Devices, Renmin University of China, Beijing 100872, People's Republic of China}
\author{Anmin\,Zhang}
\affiliation{Department of Physics, Beijing Key Laboratory of Opto-Electronic Functional Materials and Micro-nano Devices, Renmin University of China, Beijing 100872, People's Republic of China}
\author{Jianting\,Ji}
\affiliation{Department of Physics, Beijing Key Laboratory of Opto-Electronic Functional Materials and Micro-nano Devices, Renmin University of China, Beijing 100872, People's Republic of China}
\author{Yong\,Tian}
\affiliation{Department of Physics, Beijing Key Laboratory of Opto-Electronic Functional Materials and Micro-nano Devices, Renmin University of China, Beijing 100872, People's Republic of China}
\author{Kai\,Liu}
\affiliation{Department of Physics, Beijing Key Laboratory of Opto-Electronic Functional Materials and Micro-nano Devices, Renmin University of China, Beijing 100872, People's Republic of China}
\author{Youguo\,Shi}
\affiliation{Beijing National Laboratory for Condensed Matter Physics, Institute of Physics, Chinese Academy of Sciences, Beijing 100190, People's Republic of China}
\author{Tianlong\,Xia}
\affiliation{Department of Physics, Beijing Key Laboratory of Opto-Electronic Functional Materials and Micro-nano Devices, Renmin University of China, Beijing 100872, People's Republic of China}
\author{Qingming\,Zhang}
\email[Corresponding author: ]{qmzhang@ruc.edu.cn}
\affiliation{Department of Physics, Beijing Key Laboratory of Opto-Electronic Functional Materials and Micro-nano Devices, Renmin University of China, Beijing 100872, People's Republic of China}

\begin{abstract}
  We have performed polarized and temperature-dependent Raman scattering measurements on extremely large magnetoresitance compounds $\mathrm{TaAs}_{2}$ and $\mathrm{NbAs}_{2}$. In both crystals, all the Raman active modes, including six $A_g$ modes and three $B_g$ modes, are clearly observed and well assigned with the combination of symmetry analysis and first-principles calculations. The well-resolved periodic intensity modulations of the observed modes with rotating crystal orientations, verify the symmetry of each assigned mode and are fitted to experimentally determine the elements of Raman tensor matrixes. The broadening of two $A_g$ modes seen in both compounds allows us to estimate electron-phonon coupling constant, which suggests a relatively small electron-phonon coupling in the semimetals $\mathrm{TaAs}_{2}$ and $\mathrm{NbAs}_{2}$. The present study provides the fundamental lattice dynamics information on $\mathrm{TaAs}_{2}$ and $\mathrm{NbAs}_{2}$ and may shed light on the understanding of their extraordinary large magnetoresistance.
\end{abstract}


\maketitle

\section{introduction}
Magnetoresistance effect (MR) refers to the phenomenon that electrical resistance of a compound changes with the variation of applied magnetic fields. This effect has huge potential applications in magnetic storage devices, magnetic sensors and other fields, and has become one of the important research frontiers since it was proposed. The  conventional MR effects like giant magnetoresistance (GMR) and colossal magnetoresistance (CMR), were found to be negative in most cases and dominantly related to spin degrees of freedom of electrons. GMR is usually seen in the films containing magnetic ions\cite{Baibich1988, Binasch1989} and CMR appears in the manganese-based perovskites\cite{Ramirez1997a, Salamon2001}. Recent discoveries of the huge positive magnetoresistance effect in non-magnetic materials like polycrystalline silver chalcogenides ($\mathrm{Ag}_{2\mathrm{-}\mathit{\delta}}\mathrm{Te/Se}$), $\mathrm{WTe}_{2}$, $\mathrm{NbSb}_{2}$, $\mathrm{Cd}_{3}\mathrm{As}_{2}$ etc., have stimulated considerable experimental and theoretical interests\cite{Ali2014,Wang2014,Liang2015,Guo2016}. The  effect in polycrystalline silver chalcogenides was considered to be related to its linear energy dispersion at the quantum limit\cite{Xu1997,Abrikosov1998}. It was proposed that the large parabolic-field-dependent MR in $\mathrm{WTe}_{2}$\cite{Ali2014} and $\mathrm{NbSb}_{2}$\cite{Wang2014} is caused by the perfect electron-hole compensation. And the similar understanding has also been applied to bismuth\cite{Fauque2009,Yang2000}. This picture was supported by ARPES\cite{Pletikosi2014} and quantum oscillation experiments\cite{Zhu2015}. The linear-field-dependence MR of $\mathrm{Cd}_{3}\mathrm{As}_{2}$ was, however, considered to originate from the recovery of backscattering which is strongly suppressed in zero magnetic field\cite{Liang2015}.
\par
For WTe$_2$, the ultrafast carrier dynamics experiments\cite{Dai2015} show that the phonon-assisted electron-hole recombination, which is dominated by the interband electron-phonon scattering, plays a key role in boosting the large MR. And for the conventional CMR materials like La$_{1-x}$Sr$_x$MnO$_3$, it has also been proposed\cite{Millis1996,Millis1998} that the polaron effect induced by strong electron-phonon coupling (EPC), is essential to understand the CMR effect. The above examples in different systems suggest that EPC may be an important factor affecting large/colossal MR.

Recently, the semimetals $\mathrm{TaAs}_{2}$ and $\mathrm{NbAs}_{2}$ were discovered and reported to exhibit both giant positive magnetoresistance \cite{Wu2016,Wang2016,Yuan2016} and negative longitudinal magnetoresistance\cite{Luo2016} at low temperatures. The novel MR effect immediately attracted much attention in this field and the origin of it is still a puzzle. So far, no Raman study has been carried out in the newly synthesized compounds $\mathrm{TaAs}_{2}$ and $\mathrm{NbAs}_{2}$. The information on lattice dynamics and EPC in these two compounds is highly required since it is crucial to the understanding of their mechanical, thermodynamics and electronic properties, and may shed light on the  mechanism of the observed extremely large magnetoresistance.

\begin{figure*}[t]
\centering
\scalebox{0.65}{\includegraphics{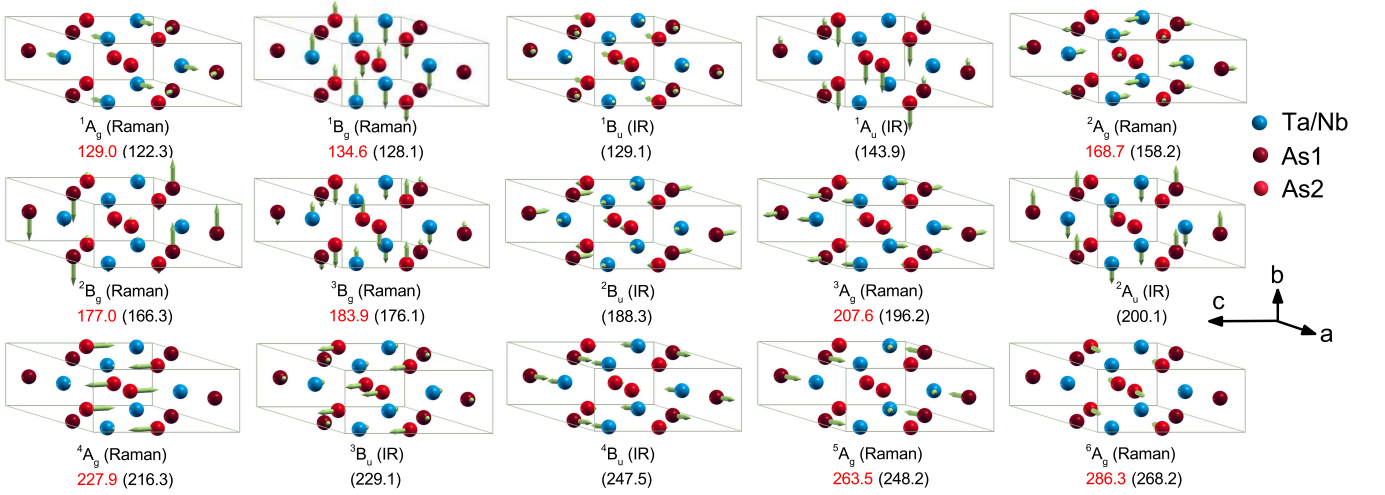}}
 \caption{\label{fig1}
Displacement patterns for all the optical modes of $\mathrm{TaAs}_{2}$ and $\mathrm{NbAs}_{2}$. The corresponding irreducible representation is indicated below each vibrational pattern. The optical activity (IR = infrared active), the experimental (red) and calculated (black) mode frequencies for $\mathrm{TaAs}_{2}$ are also listed below. The atomic structures and vibrational displacement patterns were prepared with the XCRYSDEN program\cite{Kokalj2003}.}
\end{figure*}

\par
In this paper, we have conducted polarized Raman measurements of $\mathrm{TaAs}_{2}$ and $\mathrm{NbAs}_{2}$ single crystals. All the Raman active phonon modes, six $A_g$ modes and three $B_g$ modes at Brillouin zone center, were observed. The observed Raman modes are well assigned with the combination of careful symmetry analysis and first-principles calculations. The angle dependence of phonon intensities with rotating crystal orientations, verifies the symmetry of each mode and allows to derive the elements of Raman tensor matrixes. We have further performed the temperature-dependent Raman measurements, which indicate that the temperature dependence of all the Raman modes can be well described by conventional inharmonic phonon decay process. And among the modes, two $A_g$ modes exhibit a relatively large broadening, which allows to estimate electron-phonon coupling constant and suggests a small electron-phonon coupling in both compounds.

\section{Experiments and methods}

\begin{figure*}
\scalebox{0.6}{\includegraphics*{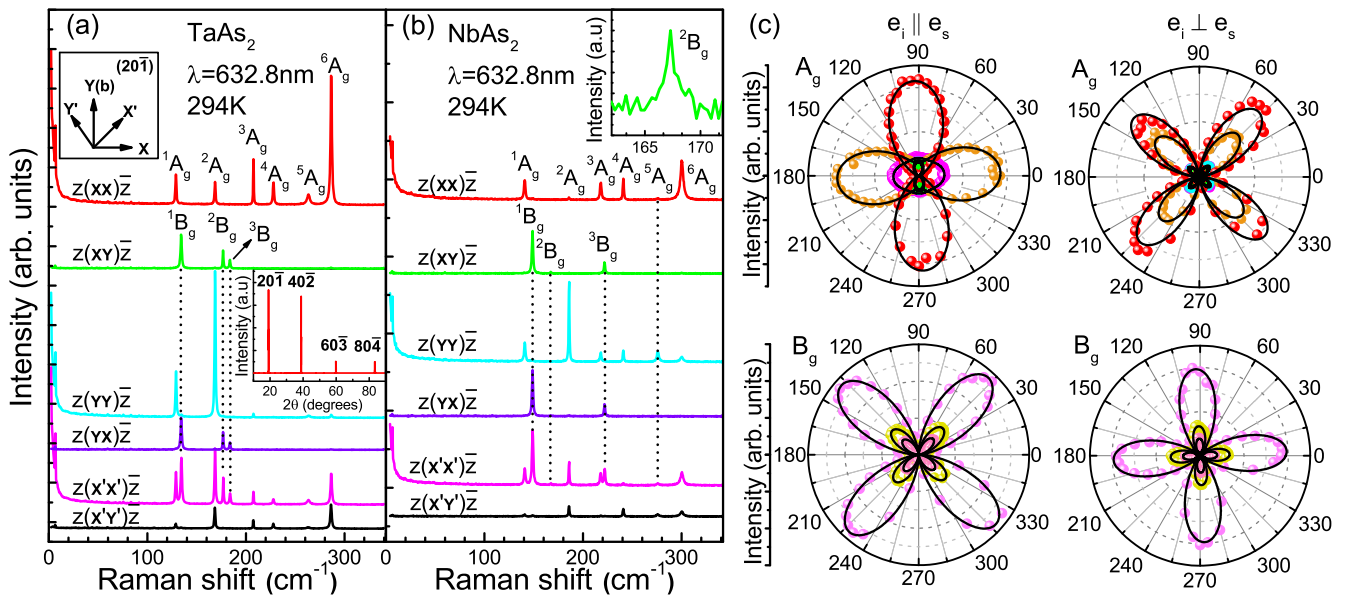}}
\caption{\label{fig2}
(a) Polarized Raman spectra of $\mathrm{TaAs}_{2}$ collected on (20$\bar{1}$) plane
at room temperature. Upper inset: the defined directions used in the measurements. Lower inset: XRD data showing high quality of single crystal. (b) Polarized Raman spectra of $\mathrm{NbAs}_{2}$ at room temperature. The low-intensity $^2B_g$ mode is zoomed in (upper inset). (c) Angle dependence (with respect to b axis) of  phonon integral intensities for $\mathrm{TaAs}_{2}$ with $\hat{\textbf{e}}_i\parallel \hat{\textbf{e}}_s$  and $\hat{\textbf{e}}_i\perp \hat{\textbf{e}}_s$ on the (20$\overline{1}$) plane at room temperature. The colored solid circles are experimental data points and the black lines are fitting curves using the angle dependence shown in Table~\ref{table1}.}
\end{figure*}

$\mathrm{TaAs}_{2}$ and $\mathrm{NbAs}_{2}$ single crystals used in this study were grown by chemical vapor transport and carefully characterized by x-ray diffraction (XRD)\cite{Wu2016,Wang2016}. By dissociating $\mathrm{TaAs}_{2}$ and $\mathrm{NbAs}_{2}$ single crystals, we obtained pieces of glossy samples with flat surface and then quickly transferred one piece into a UHV cryostat with a vacuum of better than 10$^{-8}$ mbar. Raman spectra were collected with a LABRAM HR800 system, which is equipped with a single grating of 800 mm focus length and liquid-nitrogen-cooled CCD. About 1 mW of laser power at 632.8 nm was focused into a spot with a diameter of $\sim$5 $\mu$m on the sample surface. All our Raman measurements were performed on this flat surface, which was determined to be ($20\bar{1}$) plane by XRD method, as shown in the lower inset of Fig.~\ref{fig2}a. The angle dependence of Raman intensity was measured by fixing the polarization direction of the incident and scattered light and rotating crystal orientation with an angle error of less than 2$^{\circ}$. In this article, x and y are defined as the direction perpendicular to the b axis in the ($20\bar{1}$) plane and the direction along the b axis, respectively, while x$^\prime$ and y$^\prime$ are along 45$^{\circ}$ directions with respect to x and y. The z direction is perpendicular to the xy plane, as shown in the upper inset of Fig.~\ref{fig2}a.
\begin{table}[b]
\caption{\label{table1} Symmetry analysis for $\mathrm{TaAs}_{2}$ and $\mathrm{NbAs}_{2}$ (Space group $C2/m$, No.12). The angle dependence of $A_g$ and $B_g$ mode intensities with rotating crystals in the parallel and cross polarization configurations are also given here.}
\begin{ruledtabular}
\begin{tabular}{cccccc}

   \textrm{~}    & \textrm{Wyckoff } & \textrm{Site}     & \multicolumn{3}{c}{\textrm{$\Gamma$-point}}\\
   \textrm{Atom} & \textrm{site}     & \textrm{symmetry} & \multicolumn{3}{c}{\textrm{phonon modes}}\\\hline
   Ta/Nb  & 4\,i & $C_s$ & \multicolumn{3}{c}{$2A_g+A_u+B_g+2B_u$}\\
   As1    & 4\,i & $C_s$ & \multicolumn{3}{c}{$2A_g+A_u+B_g+2B_u$}\\
   As2    & 4\,i & $C_s$ & \multicolumn{3}{c}{$2A_g+A_u+B_g+2B_u$}\\
   \multicolumn{6}{c}{Modes classification}\\
   \multicolumn{6}{l}{$\Gamma_{R}=6A_g+3B_g, \Gamma_{IR}=2A_u+4B_u, \Gamma_{acoustic}=A_u+2B_u$}\\\\
   \multicolumn{2}{c}{Scattering}    & \multicolumn{4}{c}{Angle dependence of Raman intensities}\\
   \multicolumn{2}{c}{configuration} & \multicolumn{2}{c}{$A_g$} & \multicolumn{2}{c}{$B_g$}\\\hline

   \multicolumn{2}{c}{$\hat{\textbf{e}}_i\parallel \hat{\textbf{e}}_s$} & \multicolumn{2}{c}{$\mid a\sin^2(\theta)+b\cos^2(\theta)\mid^2$} & \multicolumn{2}{c}{$\mid e\sin(2\theta)\mid^2$}\\
   \multicolumn{2}{c}{$\hat{\textbf{e}}_i\perp \hat{\textbf{e}}_s$} & \multicolumn{2}{c}{$\mid (a-b)\sin(\theta)\cos(\theta)\mid ^2$} & \multicolumn{2}{c}{$\mid e\cos(2\theta)\mid^2$}\\
\end{tabular}
\end{ruledtabular}
\end{table}

\par
$\mathrm{TaAs}_{2}$ and $\mathrm{NbAs}_{2}$ are isostructural and crystallize in the monoclinic phase\cite{Wu2016,Wang2016}, with space group $C2/m$ ($C_{2h}^3$, No.12). And all the atoms occupy 4i Wyckoff positions. Nuclear site group analysis\cite{Rousseau1981} shows that all the optical vibration modes at Brillouin zone center are composed of  6$A_g$ + 3$B_g$ + 2$A_u$ + 4$B_u$, in which $A_g$ and $B_g$ are Raman (R) active, $A_u$ and $B_u$ are infrared (IR) active, as shown in Table~\ref{table1}. In order to estimate the frequencies and displacement patterns of these optical phonons, we performed first principles calculations for $\mathrm{TaAs}_{2}$ and $\mathrm{NbAs}_{2}$, in which the projector augmented wave (PAW) method\cite{Bloechl1994,*Kresse1999} as implemented in the VASP package\cite{Kresse1993,*Kresse1996a,*Kresse1996} was used to describe the core electrons. For the exchange-correlation potential, the generalized gradient approximation (GGA) of Perdew-Burke-Ernzerhof formula\cite{Perdew1996} was adopted. The kinetic energy cutoff of the plane-wave basis was set to be 300 eV. The simulations were carried out with a triclinic cell containing 2 Ta/Nb atoms and 4 As atoms. An $8 \times 8 \times 6~k$-point mesh for the Brillouin zone sampling and the Gaussian smearing with a width of 0.05 eV around the Fermi surface were employed. In structure optimization, both cell parameters and internal atomic positions were allowed to relax until all forces were smaller than 0.001eV/$\mathring{A}$. When the equilibrium structure was obtained, the phonon modes at Brillouin zone center were calculated by using the dynamic matrix method. The calculations with 6-atom cell give 15 optical modes. However, to illustrate the displacement patterns of phonon modes, we show our results in the 12-atom supercell (Fig.~\ref{fig1}) deduced from the real-space translational invariance of 6-atom cell as in Ref.~\onlinecite{zhang2012}. Below each mode we showed the corresponding irreducible representation, optical activity, as well as the experimental (calculated) phonon frequencies for $\mathrm{TaAs}_{2}$. The experimental and calculated phonon frequencies for $\mathrm{NbAs}_{2}$ are listed in Table~\ref{table2}. Among these modes, $A_g$ and $B_u$ modes are related to the vibrations in the ac plane, and $B_g$ and $A_u$ modes are the vibrations along the b axis.

\begin{table}[b]
\caption{\label{table2} Comparison of the calculated and experimental phonon energies at 10 K for $\mathrm{TaAs}_{2}$ and $\mathrm{NbAs}_{2}$. The main atom displacements of the modes are also given. R and IR refer to Raman and infrared activity, respectively.}
\begin{ruledtabular}
\begin{tabular}{cccc|cc}
   \multirow{2}{*}{\textrm{Mode}}&\multirow{2}{*}{\textrm{Main atoms}}&\multicolumn{2}{c|}{$\mathrm{TaAs}_{2}$} & \multicolumn{2}{c}{$\mathrm{NbAs}_{2}$}\\\cline{3-6}
     &   & \textrm{Calc.} & \textrm{Expt.} & \textrm{Calc.} & \textrm{Expt.}  \\
   $^1\!{A_g}$ (R) & Ta/Nb(ac)         &122.3  &129.0   &134.8  &140.8 \\
   $^2\!A_g$ (R)   & Ta/Nb,As1(ac)     &158.2  &168.7   &177.9  &185.6 \\
   $^3\!A_g$ (R)   & Ta/Nb, As1(ac)    &196.2  &207.6   &206.7  &217.8 \\
   $^4\!A_g$ (R)   & As2(ac)           &216.3  &227.9   &228.4  &240.6 \\
   $^5\!A_g$ (R)   & As1(ac)           &248.2  &263.5   &259.3  &275.6 \\
   $^6\!A_g$ (R)   & As2(ac)           &268.2  &286.3   &281.5  &300.0 \\
   $^1\!B_g$ (R)   & Ta/Nb, As2(b)     &128.1  &134.6   &141.9  &148.7 \\
   $^2\!B_g$ (R)   & As1(b)            &166.3  &177.0   &156.9  &167.3 \\
   $^3\!B_g$ (R)   & Ta/Nb, As2(b)     &176.1  &183.9   &213.2  &221.7 \\
   $^1\!A_u$ (IR)  & As2(b)            &143.9  &~       &139.9  &~ \\
   $^2\!A_u$ (IR)  & Ta/Nb(-b), As1(b) &200.1  &~       &228.3  &~ \\
   $^1\!B_u$ (IR)  & As2(ac)           &129.1  &~       &126.4  &~ \\
   $^2\!B_u$ (IR)  & As1(ac)           &188.3  &~       &183.9  &~ \\
   $^3\!B_u$ (IR)  & As2(ac)           &229.1  &~       &257.9  &~ \\
   $^4\!B_u$ (IR)  & Ta/Nb, As1(ac)    &247.5  &~       &280.3  &~ \\

\end{tabular}
\end{ruledtabular}
\end{table}

\section{Polarized and angle-dependent spectra}

Raman spectra of $\mathrm{TaAs}_{2}$ and $\mathrm{NbAs}_{2}$ collected at room temperature are shown in Fig.~\ref{fig2}(a) and \ref{fig2}(b), respectively. Experimentally, Raman intensity clearly depends on the polarizations of incident and scattered light and can be expressed as
\begin{eqnarray}
 I \propto |\hat{\textbf{e}}_i\cdot\Re\cdot\hat{\textbf{e}}_s|^2
\end{eqnarray}

where $I$ is Raman intensity, $\hat{\textbf{e}}_i$ and $\hat{\textbf{e}}_s$ are the unit vectors of the polarizations of incident and scattered light, respectively, and $\Re$ refers to Raman scattering tensor, a 3 $\times$ 3 matrix determined by symmetry of phonon mode. For the orthogonal coordinate system, Raman scattering tensor\cite{Loudon1964} of $A_g$ and $B_g$ phonons in $C_{2h}$ point group can be expressed as
\begin{center}
$A_{g}=\left(
     \begin{array}{ccc}
       a&~~0~~&d\\
       0&~~b~~&0\\
       d&~~0~~&c\\
     \end{array}
     \right),~~~$
$B_{g}=\left(
     \begin{array}{ccc}
       0&~~e~~&0\\
       e&~~0~~&f\\
       0&~~f~~&0\\
     \end{array}
     \right).$
 \end{center}

 If $\theta$ is defined as the angle between the b axis of crystal and the polarization of incident light, the intensities of $A_g$ and $B_g$ modes will vary as a function of the $\theta$ angle. Under different polarization configurations ($\hat{\textbf{e}}_i\parallel \hat{\textbf{e}}_s$, $\hat{\textbf{e}}_i\perp \hat{\textbf{e}}_s$), the angle dependence will be different and the details are shown in the bottom of Table~\ref{table1}. Raman tensors determine that only $A_g$ phonon can be observed at z(xx)$\overline{\text{z}}$ configuration ($\hat{\textbf{e}}_i\parallel \hat{\textbf{e}}_s$, $\theta$=90$^{\circ}$) and z(yy)$\overline{\text{z}}$ configuration ($\hat{\textbf{e}}_i\parallel \hat{\textbf{e}}_s$, $\theta$=0$^{\circ}$). Under the two configurations, we observed six sharp peaks for $\mathrm{TaAs}_{2}$ located at 129 ($^1A_g$), 168.7 ($^2A_g$), 207.6 ($^3A_g$),  227.9 ($^4A_g$), 263.5 ($^5A_g$), 286.3 ($^6A_g$) $\mathrm{cm}^{-1}$ (Fig.~\ref{fig2}(a)), respectively. Similarly, for $\mathrm{NbAs}_{2}$ the six peaks are located at 140.8 ($^1A_g$), 185.6 ($^2A_g$), 217.8 ($^3A_g$), 240.6 ($^4A_g$), 275.6 ($^5A_g$), 300 ($^6A_g$) $\mathrm{cm}^{-1}$ (Fig.~\ref{fig2}(b)). The frequencies of the observed phonons are all slightly larger than the calculated ones (Table~\ref{table2}). The difference in intensity between the two configurations indicates that matrix elements of Raman tensors are distinguishable for different modes. It can be expected that only $B_g$ phonons should be observed under z(xy)$\overline{\text{z}}$ ($\hat{\textbf{e}}_i\perp \hat{\textbf{e}}_s$, $\theta$=90$^{\circ}$) and z(yx)$\overline{\text{z}}$ configurations ($\hat{\textbf{e}}_i\perp \hat{\textbf{e}}_s$, $\theta$=0$^{\circ}$). Under both configurations, there are three peaks in $\mathrm{TaAs}_{2}$ located at 134.6 ($^1B_g$), 177 ($^2B_g$), 183.9 ($^3B_g$) $\mathrm{cm}^{-1}$, and the three modes in $\mathrm{NbAs}_{2}$  are observed at 148.7 ($^1B_g$), 167.3 ($^2B_g$), 221.7 ($^3B_g$) $\mathrm{cm}^{-1}$. Like $A_g$ modes, the energies of these modes are also slightly larger than the calculated values. When the polarization of the incident light and scattered light $\hat{\textbf{e}}_i$, $\hat{\textbf{e}}_s$ are not exactly perpendicular or parallel to the b axis, all Raman active phonons with different symmetries can be detected. For instance, all the $A_g$ and $B_g$ phonon modes are seen in z(x$^\prime$x$^\prime$)$\overline{\text{z}}$, z(x$^\prime$y$^\prime$)$\overline{\text{z}}$ configurations.

\begin{table}[t]
\caption{\label{table3} Fitting parameters of Raman tensors normalized by \textit{b} ($\simeq$1.485) of $^4A_g$ mode in $\mathrm{TaAs}_{2}$ and by \textit{a} ($\simeq$1.35) of $^5A_g$ mode in $\mathrm{NbAs}_{2}$. }
\begin{ruledtabular}
\begin{tabular}{cccc||ccc}
\multicolumn{4}{c||}{$\mathrm{TaAs}_{2}$} & \multicolumn{3}{c}{$\mathrm{NbAs}_{2}$}\\\hline
  ~ & a & b & e  & a & b & e  \\
  \textrm{$^1\!A_g$} & 25.90 & 31.82 & ~     & 11.869 & 11.615      & ~       \\
  \textrm{$^2\!A_g$} & 16.97 & 56.80 & ~     & 1.689  & 24.881      & ~       \\
  \textrm{$^3\!A_g$} & 25.05 & 3.05  & ~     & 11.519 & 8.363  & ~       \\
  \textrm{$^4\!A_g$} & 23.18 & 1     & ~     & 12.385 & -7.333 & ~       \\
  \textrm{$^5\!A_g$} & 22.76 & 8.06  & ~     & 1      & -8.570 & ~       \\
  \textrm{$^6\!A_g$} & 61.01 & 3.52  & ~     & 16.622 & 5.178  & ~       \\
  \textrm{$^1\!B_g$} & ~     & ~     & 32.69 & ~      & ~      & 17.496  \\
  \textrm{$^2\!B_g$} & ~     & ~     & 19.19 & ~      & ~      & 2.830   \\
  \textrm{$^3\!B_g$} & ~     & ~     & 14.68 & ~      & ~      & 9.141   \\

\end{tabular}
\end{ruledtabular}
\end{table}

\begin{table}[b]
\caption{\label{table4} Fitting parameters of the positions and linewidths of the observed modes in $\mathrm{TaAs}_{2}$ and $\mathrm{NbAs}_{2}$. The unit of $\omega(0)$ and $\Gamma_0$ is $\mathrm{cm}^{-1}$, and the unit of $\gamma$A is $10^{-7} K^{-2}$.}
\begin{ruledtabular}
\begin{tabular}{ccccc||cccc}
\multicolumn{5}{c||}{ $\mathrm{TaAs}_{2}$} & \multicolumn{4}{c}{$\mathrm{NbAs}_{2}$}\\\hline
  ~ & $\omega(0)$    & $\gamma$A   & $\Gamma_0$ & $\lambda_{ph-ph}$ & $\omega(0)$ & $\gamma$A  & $\Gamma_0$ & $\lambda_{ph-ph}$ \\
  \textrm{$^1\!A_g$}  & 130.2              & 2.46       & 0.21     	  & 0.30                   & 142.5     & 2.96		 & 0.29       & 0.37\\
  \textrm{$^2\!A_g$}  & 170.7              & 2.86       & 0.22      	  & 0.39  			    & 188.0     & 3.06 		 & 0.13       & 0.56  \\
  \textrm{$^3\!A_g$}  & 209.5    	     & 2.20 	 & 0.12   		  & 0.52  			    & 220.3     & 2.76 		 & 0.24       & 0.86 \\
  \textrm{$^4\!A_g$}  & 229.8   		     & 2.08	      & 0.17   		  & 0.92 			    & 243.8     & 3.16		      & 0.17       & 0.74 \\
  \textrm{$^5\!A_g$}  & 266.1    	     & 2.36	      & 0.51   	       & 1.00 			    & 279.0     & 2.98		      & 0.38       & 1.02 \\
  \textrm{$^6\!A_g$}  & 289.7    	     & 2.78	      & 0.25            & 1.05  			   & 304.3      & 3.48 		 & 0.54       & 1.31 \\
  \textrm{$^1\!B_g$}  & 135.9   	 	     & 2.60 	 & 0.11    	       & 1.03 			   & 150.4      & 2.66 		 & 0.18       & 0.66 \\
  \textrm{$^2\!B_g$}  & 178.3    	     & 1.90 	 & 0.15    	       & 0.30  			   & 168.9      & 2.32 		 & 0.13       & 1.39\\
  \textrm{$^3\!B_g$}  & 185.6  	          & 2.24 	 & 0.07           & 1.23   		  	   & 224.7      & 3.18 		 & 0.19       & 0.85  \\
\end{tabular}
\end{ruledtabular}
\end{table}

 \par
We further made angle-dependent measurements to verify the symmetries of the observed Raman modes by rotating crystal orientation around the normal direction of ($20\bar{1}$) plane. In Fig.~\ref{fig2}(c), we showed the angle dependence of the integrated intensities of all the nine modes under the two configurations, $\hat{\textbf{e}}_i\parallel \hat{\textbf{e}}_s$ and $\hat{\textbf{e}}_i\perp \hat{\textbf{e}}_s$, for $\mathrm{TaAs}_{2}$. For $\hat{\textbf{e}}_i\parallel \hat{\textbf{e}}_s$ configuration, the six $A_g$ modes have twofold symmetry and the three $B_g$ modes have fourfold symmetry featured by a node at 0$^{\circ}$ and a maximum at 45$^{\circ}$. Under $\hat{\textbf{e}}_i\perp \hat{\textbf{e}}_s$ configuration, the symmetry of all the modes are fourfold and the six $A_g$ modes are characterized by a node at 0$^{\circ}$ and a maximum at 45$^{\circ}$, while the three $B_g$ ones are rotated by 45$^{\circ}$. The experimental angle-dependent data can be well fitted using the formula in Table~\ref{table1} (solid curves in Fig.~\ref{fig2}(c)). The fitting parameters are summarized in Table~\ref{table3}, which reflect the matrix elements of Raman tensors. It should be noted that the experimental maximum of phonon intensity are shifted from the expected one by a few degrees. This may originate from a small deviation of the angle between $\hat{\textbf{e}}_i$ and $\hat{\textbf{e}}_s$ from the right angle. The intensity oscillations of $^1A_g$ and $^2A_g$ modes are in anti-phase with those of $^3A_g$, $^4A_g$, $^5A_g$, $^6A_g$ modes in the $\hat{\textbf{e}}_i\parallel \hat{\textbf{e}}_s$ configuration.  This is due to the different relative magnitudes of matrix elements of Raman tensors, a and b. All the angle-dependent observations are well consistent with the above assignment.

\section{Temperature-dependent spectra}
\begin{figure*}
\scalebox{0.43}{\includegraphics*{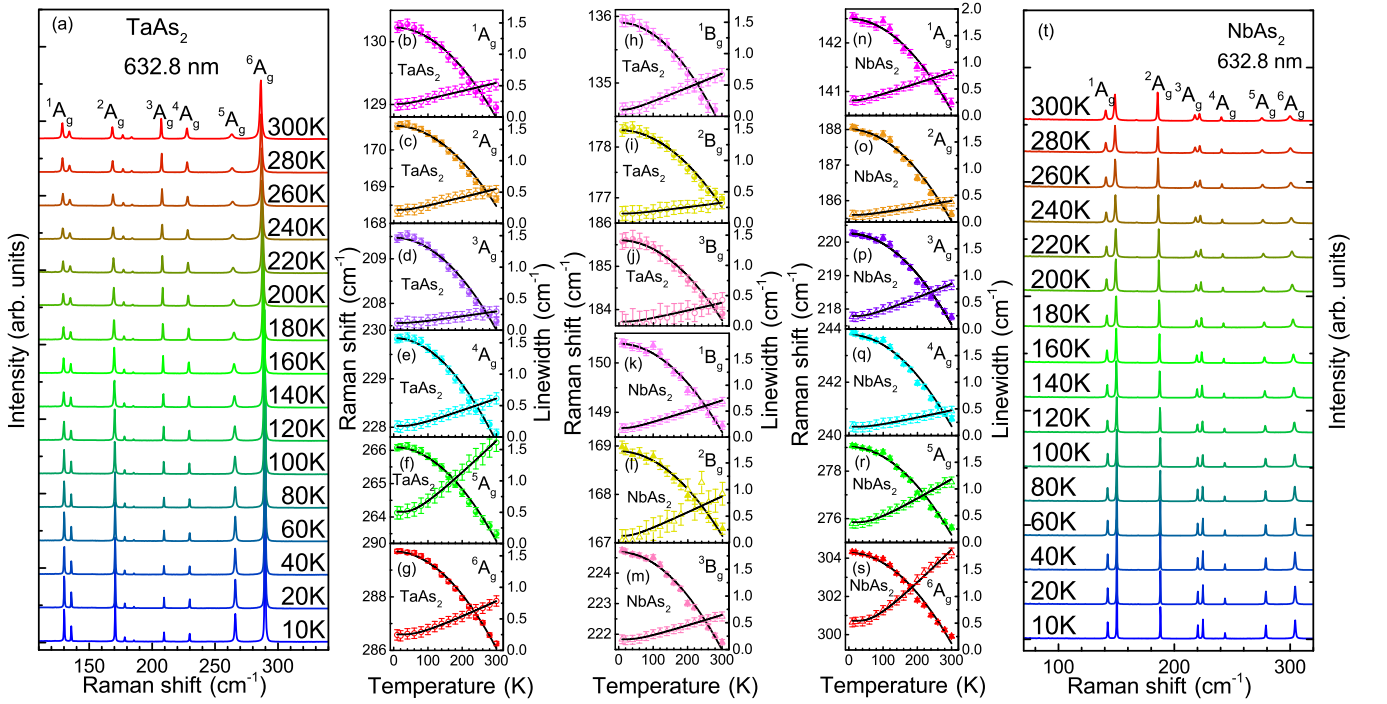}}
\caption{\label{fig3}  Temperature dependence of Raman spectra of $\mathrm{TaAs}_{2}$ (a) and $\mathrm{NbAs}_{2}$ (t) from 10 to 300 K. Temperature dependence of the extracted positions and linewidths (HWHM) of the modes in $\mathrm{TaAs}_{2}$ ((b)-(j)) and $\mathrm{NbAs}_{2}$ ((k)-(s)). The data points in (b)-(s) are extracted from Raman spectra in (a) and (f) by Voigt function fitting. The solid lines in (b)-(s) are the fitting curves using Eqs.~(\ref{peaktemdep}) and Eqs.~(\ref{HWHMtemdep}) (see text). }
\end{figure*}

Temperature-dependent Raman scattering measurements on the two compounds were also carried out from 10 to 300 K. Fig.~\ref{fig3}(a) and \ref{fig3}(t) display the temperature-dependent spectra of $\mathrm{TaAs}_{2}$ and $\mathrm{NbAs}_{2}$, respectively. It is clear that all the phonon peaks become sharper with cooling and the symmetric line shapes suggest the absence of strong EPC in the two materials.

In Fig.~\ref{fig3}(b)-(j) and ~\ref{fig3}(k)-(s), we show the temperature dependence of peak positions of the nine modes of $\mathrm{TaAs}_{2}$ and $\mathrm{NbAs}_{2}$, which have been fitted with Voigt functions. In general, the temperature dependence of phonon frequencies $\omega_i(T)$ is mainly contributed by the lattice thermal expansion and anharmonic interaction, and can be expressed\cite{Eiter2014,Opacic2015} as:
\begin{eqnarray}
\label{peaktemdep}
\omega_i(T)=\omega_i(0)+\Delta_i^V(T)+\Delta_i^A(T),
\end{eqnarray}
where the $\omega_i(0)$ is the harmonic frequency of an optical mode at zero temperature. The second term in (\ref{peaktemdep}) describes the contribution of the lattice thermal expansion, and can be written as
\begin{eqnarray}
\Delta_i^V(T)=\omega_i(0)\left(e^{-\gamma_i\int_0^T\alpha(T')dT'}-1\right),
\end{eqnarray}
where $\gamma_i$ is the Gr\"uneisen parameter and $\alpha(T)$ is the thermal expansion coefficient.

The third term in (\ref{peaktemdep}) represents the change in phonon energy due to the anharmonic interaction with other phonons. If the anharmonic effect is described by three-phonon processes, it follows that 
\begin{eqnarray}
\Delta_i^A(T)=-\frac{2\Gamma_{0,i}^2}{\omega_i(0)}\left(1+\frac{4\lambda_{ph-ph,i}}{e^{\hbar\omega_i(0)/2k_\emph{B}T}-1}\right),
\end{eqnarray}
The width $\Gamma_{0,i}$ of the $i$th Raman line can be obtained by $\Gamma_i(T)$ extrapolated to zero temperature. And $\lambda_{ph-ph,i}$ represents the phonon-phonon coupling strength.

To the best of our knowledge, the experimental Gr\"uneisen parameter $\gamma_i$ and thermal expansion coefficient $\alpha(T)$ of $\mathrm{TaAs}_{2}$ and $\mathrm{NbAs}_{2}$ are still lacking so far. For simplicity, the thermal expansion coefficient $\alpha(T)$ is assumed to be roughly linear with temperature, i.e., $\alpha(T) = AT$. And the coefficient A and Gr\"uneisen parameter $\gamma_i$ are combined together as a single fitting parameter (see Table~\ref{table4}). The treatment allows a very good fit to our data and the fitting curves are displayed in Fig.~\ref{fig3}.
\par
We also display the temperature dependence of linewidths (Half Width at Half Maximum, HWHM) of the nine modes in Fig.~\ref{fig3}. It should be noted that the linewidths of As-related modes, for example, $^5A_g$, are greatly narrow than those in other materials\cite{Rahlenbeck2009}. On one hand, this reflects the high quality of the crystals used in our measurements. On the other hand, it may also be an indication of small electron-phonon coupling interaction in the two compounds. The temperature dependence of phonon linewidths (Fig.~\ref{fig3}) can be described by multi-phonon decay process \cite{Klemens1966}, in which an optical phonon with zero wave vector and a finite frequency $\omega$, decays into two acoustic phonons with opposite wave vectors and equal frequencies $\sim\omega/2$. And the corresponding expression \cite{Eiter2014} is written as
\begin{eqnarray}
\label{HWHMtemdep}
\Gamma_i(T)=\Gamma_{0,i}\left(1+\frac{2\lambda_{ph-ph,i}}{e^{\hbar\omega_i(0)/2k_\emph{B}T}-1}\right).
\end{eqnarray}

Interestingly, the linewidths of $^5A_g$ and $^6A_g$ modes in both samples extrapolated to zero temperature, are almost two times larger than those of other $A_g$ modes. The phonon-phonon scattering at zero temperature is negligible due to phonon freezing, and phonon linewidths at zero temperature mainly comes from impurities and electron-phonon scattering. If the contribution from impurities scattering is approximately equal for all the phonon modes, the larger linewidths of $^5A_g$ and $^6A_g$ modes suggest a finite electron-phonon coupling in the compounds.

\par
The averaged linewidths of  $^1A_g$ to $^4A_g$ phonon modes extrapolated to zero temperature, can be approximately taken as the contributions caused by impurities. Then the linewidths ($\Gamma_{e-ph}$) contributed by electron-phonon scattering for $^5A_g$ and $^6A_g$, are 0.33($\pm$0.099) $\mathrm{cm}^{-1}$, 0.07($\pm$0.078)  $\mathrm{cm}^{-1}$ in $\mathrm{TaAs}_{2}$, and 0.17($\pm$0.083)  $\mathrm{cm}^{-1}$ and 0.34($\pm$0.081)  $\mathrm{cm}^{-1}$ in $\mathrm{NbAs}_{2}$, respectively. Generally, electron-phonon coupling constant associated with a particular mode can be estimated by Allen formula\cite{Allen1974}:
\begin{eqnarray}
\lambda_{e-ph} = 2\times\Gamma_{e-ph} / (\pi N_{E_f} \hbar\omega_i(0)^2 ).
\end{eqnarray}
where $N_{E_f}$ is electronic density of states at Fermi level per eV per spin per unit cell, $\omega_i(0)$ the bare frequency in the absence of electron-phonon coupling. For $^5A_g$ mode of $\mathrm{TaAs}_{2}$, $\Gamma_{e-ph}$ = 0.33 $\mathrm{cm}^{-1}$, $\omega_i(0)$ = 267.4 $\mathrm{cm}^{-1}$, and $N_{E_f}$ = 0.76 states/eV/spin/unit cell from our band structure calculations, we have $\lambda_{e-ph}$ = 0.031($\pm$0.009). Similarly, $\lambda_{e-ph}$ is estimated to be 0.006($\pm$0.006) for $^6A_g$ mode in $\mathrm{TaAs}_{2}$. Applying the same procedure to $^5A_g$ and $^6A_g$ modes of $\mathrm{NbAs}_{2}$, we obtained electron-phonon coupling constants 0.014($\pm$0.007) and 0.023($\pm$0.005), respectively. It should be pointed out that these values seem quite close to that  of $\mathrm{WTe}_{2}$ ($\lambda_{e-ph}\sim$0.016)\cite{mynote}. The small electron-phonon coupling constants indicate that electron-phonon coupling may play a relatively small role in this kind of large magnetoresitance semimetal materials. The situation is completely different in the manganese-based perovskites CMR materials, where electron-phonon coupling is rather strong ($\sim$1) and is important to the understanding of its colossal magnetoresistance\cite{Millis1996,Millis1998}. On the other hand, EPC seems to be involved into the large MR in some way. The ultrafast carrier-dynamics experiments\cite{Dai2015} in $\mathrm{WTe}_{2}$ show that the phonon-assisted electron-hole recombination, which is dominated by the interband electron-phonon scattering, is possibly helpful to the enhancement of the large MR. Actually the present study raises a fundamental issue: what role EPC plays in the large MR?  A comprehensive theoretical framework and more experiments are needed to address the issue.

\section{Summary}
In summary, we have carried out polarized and temperature-dependent Raman measurements on $\mathrm{TaAs}_{2}$ and $\mathrm{NbAs}_{2}$ single crystals. All the Raman active phonons are observed and well assigned through a careful symmetry analysis and first-principles calculations. The angle dependence of phonon intensities verifies the symmetry for each assigned mode and further gives the elements of Raman tensor matrixes. The small electron-phonon coupling constants, which are derived from the broadening of two $A_g$ phonons, indicate that electron-phonon interaction may play a relatively small role in understanding the large magnetoresitance in the semimetal materials.

\begin{acknowledgments}
This work was supported by the Ministry of Science and Technology of China (Grant No.: 2016YFA0300504) and the NSF of China. Q.M.Z., K.L. and T.L.X. were supported by the Fundamental Research Funds for the Central Universities and the Research Funds of Renmin University of China. Y.G.S. was supported by the Strategic Priority Research Program (B) of the Chinese Academy of Sciences (Grant No. XDB07020100). Computational resources have been provided by the Physical Laboratory of High Performance Computing at Renmin University of China.
\end{acknowledgments}

\bibliography{Reference}

\end{document}